\documentclass[osajnl,reprint,letter,amsmath,amssymb]{revtex4-1}
\usepackage[utf8]{inputenc}
\usepackage[english]{babel}
\usepackage{graphicx}
\usepackage{subfig}
\usepackage{dcolumn}
\usepackage{textcomp}
\usepackage{rotating}
\newcommand{\SiM}[1]{$^{#1}\mathrm{Si}$}
\newcommand{\cmminusi}{$\textrm{cm}^{-1}$}
\newcommand{\mkm}{\textmu{}m}
\newcommand{\Celsius}{$^\circ$C}
\begin{document}
\title{%
Refractive index spectral dependence, Raman and transmission spectra \\
of high-purity {\boldmath \SiM{28}, \SiM{29}, \SiM{30}, and \SiM{\textrm{nat}}}
single crystals
}
\author{V.G.Plotnichenko}
\email[E-mail:~~]{victor@fo.gpi.ac.ru}
\author{V.O.Nazaryants}
\author{E.B.Kryukova}
\author{V.V.Koltashev}
\author{V.O.Sokolov}
\author{E.M.Dianov}
\affiliation{Fiber~Optics~Research~Center of the~Russian~Academy~of~Sciences \\
38~Vavilov~Street, Moscow 119333, Russia}
\author{A.V.Gusev}
\author{V.A.Gavva}
\author{T.V.Kotereva}
\author{M.F.Churbanov}
\affiliation{Institute of~Chemistry of~High-Purity~Substances
of the~Russian~Academy~of~Sciences \\
49~Tropinin~Street, Nizhny~Novgorod 603600, Russia}
\begin{abstract}
Precise measurement of the refractive index of stable silicon isotopes \SiM{28},
\SiM{29}, \SiM{30} single crystals  with enrichments above 99.9~at.\%{} and a
silicon single crystal \SiM{\textrm{nat}} of natural isotopic composition is
performed with the Fourier-transform interference refractometry method from 1.06
to more than 80~\mkm{} with 0.1~\cmminusi{} resolution and accuracy of
$2\times10^{-5} \ldots 1\times10^{-4}$. The oxygen and carbon concentrations in
all crystals are within $5\times10^{15}\textrm{~cm}^{-3}$ and the content of
metal impurities is $10^{-5} \ldots 10^{-6}$~at.\%. The peculiar changes of the
refractive index in the phonon absorption region of all silicon single crystals
are shown. The coefficients of generalized Cauchy dispersion function
approximating the experimental refractive index values all over the measuring
range are given. The transmission and Raman spectra are also studied.
\end{abstract}
\maketitle

\section{Introduction}
Production and study of the properties of isotopically enriched substances are
one of the crucial problems of contemporary material science involving in
itself the increasing number of researchers. Owing to the achieved level of
purity of many substances, their properties are substantially determined by the
isotopic composition.

Development of the technology of high-purity monoisotopic single crystals
production allows one to study the influence of the isotopic composition on
physical properties, in particular on the electronic, phonon, and Raman spectra
and the refractive index dispersion. This is important both for verification of
existing theoretical models and filling them with more correct values of used
parameters, and for calculation of parameters of next-generation devices
developed on their basis.

Crystalline silicon is known to be the basic material of microelectronics and
solar energetics. Besides, it is also an attractive optical material with high
transmission in the near- to mid-IR spectral range. In recent years silicon
photonics grows rapidly, which is expected to find application in future optical
fiber communication and information transfer systems. Monoisotopic silicon is
considered to be the basic component of quantum computers \cite{1}.

Formerly silicon refractive index was measured repeatedly in various spectrum
ranges by many laboratories and companies (see e.g. \cite{17, 18, 19, 20, 21}).
The papers \cite{17, 18} seem to be the first in this field. The analysis of
more than 50 papers made in 1980 \cite{19} showed the data available to be
scattered widely, but allowed to derive certain averaged spectral dependence for
silicon refractive index which is rather often referred to. Unfortunately, in
the majority of published papers devoted to silicon refractive index measuring,
the crystals used for this purpose are characterized neither in purity, nor in
conductivity, both influencing the measurement results.

In this paper we present the results of measuring the room-temperature
transmission spectra in the short-wavelength edge and phonon absorption regions,
the Raman spectra excited by the 514.5~nm Ar laser line and the refractive index
spectral dependence in the \hbox{near-}, \hbox{mid-}, and far-IR spectral ranges
in silicon single crystals with natural isotopic composition,
\SiM{\textrm{nat}}, and silicon isotopes \SiM{28}, \SiM{29}, \SiM{30}{} with the
basic isotope content above 99.9~at.\%{} prepared by the same technology. A
short note on fabrication of these single crystals and refractive index
measurements in the $1.06 \ldots 25.5$~\mkm{} range has been published recently
\cite{2}.

\section{Preparation of single crystals and samples for measurements}
Polycrystalline silicon was prepared by thermal decomposition of monosilane
enriched in one of the silicon isotopes as described in \cite{3, 4}. The
decomposition was proceed on a graphite substrate surface at 800~\Celsius. After
a few mm thick layer was deposited, polycrystalline silicon was separated from
the graphite substrate, remelted to form a rod, and purified by 10 float zoning
passes. Then this monoisotopic silicon rod was used as a substrate to deposit
the polycrystalline silicon layer till an ingot of a required diameter was
obtained.

To minimize the effect of isotopic dilution by a seeding material, a special
growth procedure \cite{5} was used. Preliminarily a single-crystal seed with
the isotopic composition gradient in length was grown by drawing at the end
of mono-isotopic silicon ingot of a cylindrical site with a diameter not
exceeding that of a seed from silicon of natural isotopic composition. The
composition of grown single crystal seed was almost identical to the isotopic
composition of polycrystalline ingot which the single crystal was grown from.
Single crystals were grown in the $\left[100\right]$ direction by float zoning
in a high-purity argon atmosphere. The same procedure was used to prepare a
single crystal of natural isotopic composition, \SiM{\textrm{nat}}: \SiM{28}{}
(92.23\%), \SiM{29}{} (4.68\%), and \SiM{30}{} (3.09\%) \cite{6}.
\begin{table}
\caption{%
Results of mass spectrometry determination of isotopic composition of
produced Si single crystals
}
\begin{tabular}{c||D{.}{.}{5}|D{.}{.}{5}|D{.}{.}{5}}
\hline
\hline
&\multicolumn{3}{c}{}                                    \\[-2.00ex]
Single & \multicolumn{3}{c}{Isotopic composition, at.\%} \\
\cline{2-4}
&&&                                                      \\[-2.00ex]
crystal &
\multicolumn{1}{c|}{\SiM{28}} &
\multicolumn{1}{c|}{\SiM{29}} &
\multicolumn{1}{c}{\SiM{30}}                             \\
\hline
\hline
&&&                                                      \\[-2.00ex]
\SiM{28} & 99.9934  &  0.00637  &  0.00023               \\
\SiM{29} &  0.026   & 99.919    &  0.055                 \\
\SiM{30} &  0.005   &  0.021    & 99.974                 \\
\hline
\hline
\end{tabular}
\label{tab:1}
\end{table}

The isotopic composition of the crystals was determined by mass spectrometry
\cite{7} with an accuracy better than 0.01~at.\%{} (Table~\ref{tab:1}).
Comparison with the data reported in \cite{8, 9} indicates that our crystals
are more isotopically enriched and offer higher chemical purity and resistivity.

Structural perfection of the crystals was evaluated using X-ray diffraction and
selective etching. No orientation disorder, low-angle boundaries or twinning
regions were detected in the single crystals grown. The half-width of their
rocking curves was $11 \ldots 12$''. The room-temperature resistivity of the
crystals was found to be 100--200~$\Omega \times \textrm{cm}$. According to
mass spectrometric measurements, the contents of 72 impurities was below the
mass spectrometry detection limit ($\lesssim 10^{-5} \ldots 10^{-6}$~at.\%)
\cite{10}.

$2 \ldots 3$~mm-thick plates were cut perpendicularly to the growth axis of
the cylindrical samples prepared. Then the plates were ground and polished
with various flatness degrees for measuring the transmission spectra without
interference at 2~\cmminusi{} resolution (to analyze the presence of impurity
absorption bands and intrinsic bands shift in the two-phonon absorption range)
and with interference at 0.1~\cmminusi{} resolution (to measure the refractive
index spectral dependence). Samples for the refractive index measurements had
the surface flatness better than $1/10$ of Newton's fringe ($\lambda = 546$~nm),
parallelism of sides better than 1\textquotesingle\textquotesingle, and the
surface finish class of 60/40.

\section{Transmission spectra}
The transmission spectra of crystalline silicon without interference were
measured on Lambda~900 spectrophotometer in a short-wavelength range
($0.9 \ldots 1.6$~\mkm) and on Bruker~IFS-113v Fourier-transform vacuum
spectrometer in a longer wavelength range ($1 \ldots 90$~\mkm). The analysis
on the  presence of impurity absorption bands in the transmission spectra
measured in conformity with the ASTM standards F-1188 and F-1391 \cite{11}
proved oxygen and carbon contents to be less than $5 \times 10^{15}$~cm$^{-3}$
in all the samples under investigation.

In the 1 to 3~mm thickness crystals the tendency to the transmission edge shift
to higher frequencies with isotope atomic mass increasing was observed in the
short-wavelength transmission edge region at room temperature. However, the
shift value of $\lesssim 1\textrm{~meV}/\textrm{a.m.u.}$ was comparable with the
experimental error.

The $M^{-1/2}$ dependence of the band gap, $E_g$, was shown in \cite{12} to be
more probable for $T < 50$~K, since due to lattice expansion and electron-phonon
interaction $E_g$ decreases linearly with $T$ growing in the room-temperature
region ($\sim 294$~K). Measurement at 16~K of the transmission spectra of our
silicon samples showed that with atomic mass increasing the absorption edge
shifts to the high-energy region. The shift was found to be $\mp \left(1.8 \pm
0.3\right)$~meV in \SiM{28}{} and \SiM{30}{} relative to \SiM{29}.
The indirect transition energy determined from the optical absorption edge was
found to depend on the average atomic mass as $M^{-1/2}$. The shift $\sim
1.5$~meV measured in the present work at 16~K turned out to be somewhat higher
compared with 1~meV derived from the luminescence spectra in \cite{12}.

In the phonon absorption region ($5 \ldots 30$~\mkm) a shift of all absorption
bands to longer wavelengths was observed, which had been seen previously in
isotopically enriched silicon single-crystal samples with a smaller isotopic
enrichment \cite{13}.

\section{Raman spectra}
The Raman spectra of all produced single crystals were measured on a T64000
triple spectrograph in a backscattering geometry at room temperature. The
514.5-nm argon laser line was used as an incident radiation. The scattered
radiation was detected by a liquid nitrogen-cooled CCD matrix. The spectra
were gathered in 0.7~\cmminusi{} steps. Spectrograph was calibrated using
radiation in Ar laser and Hg lamp lines.

\begin{figure*}
\subfloat[\label{fig:1a}]{
\includegraphics[scale=0.50,bb= 50 260 550 770]{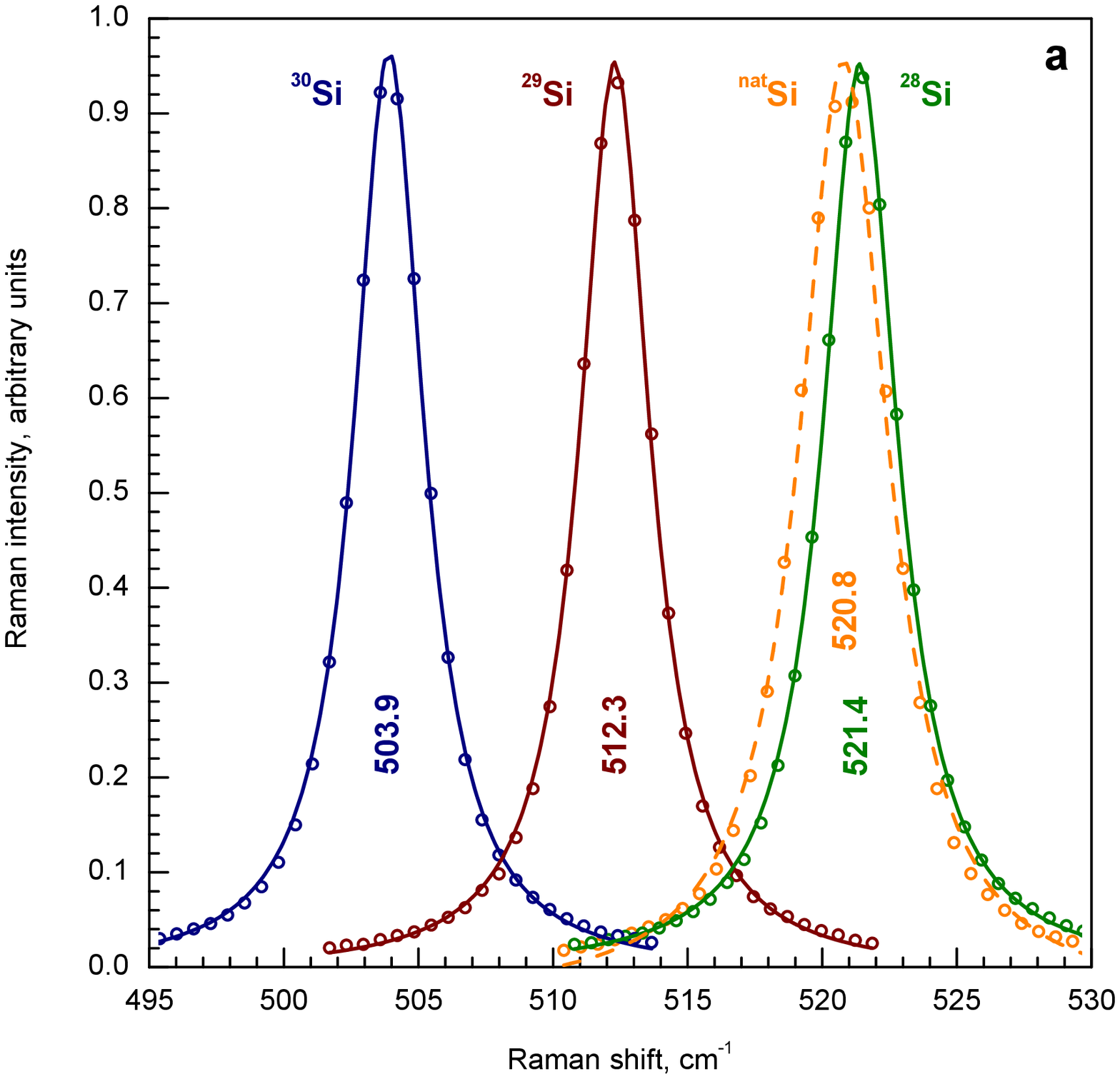}}
\subfloat[\label{fig:1b}]{
\includegraphics[scale=0.50,bb= 50 260 550 770]{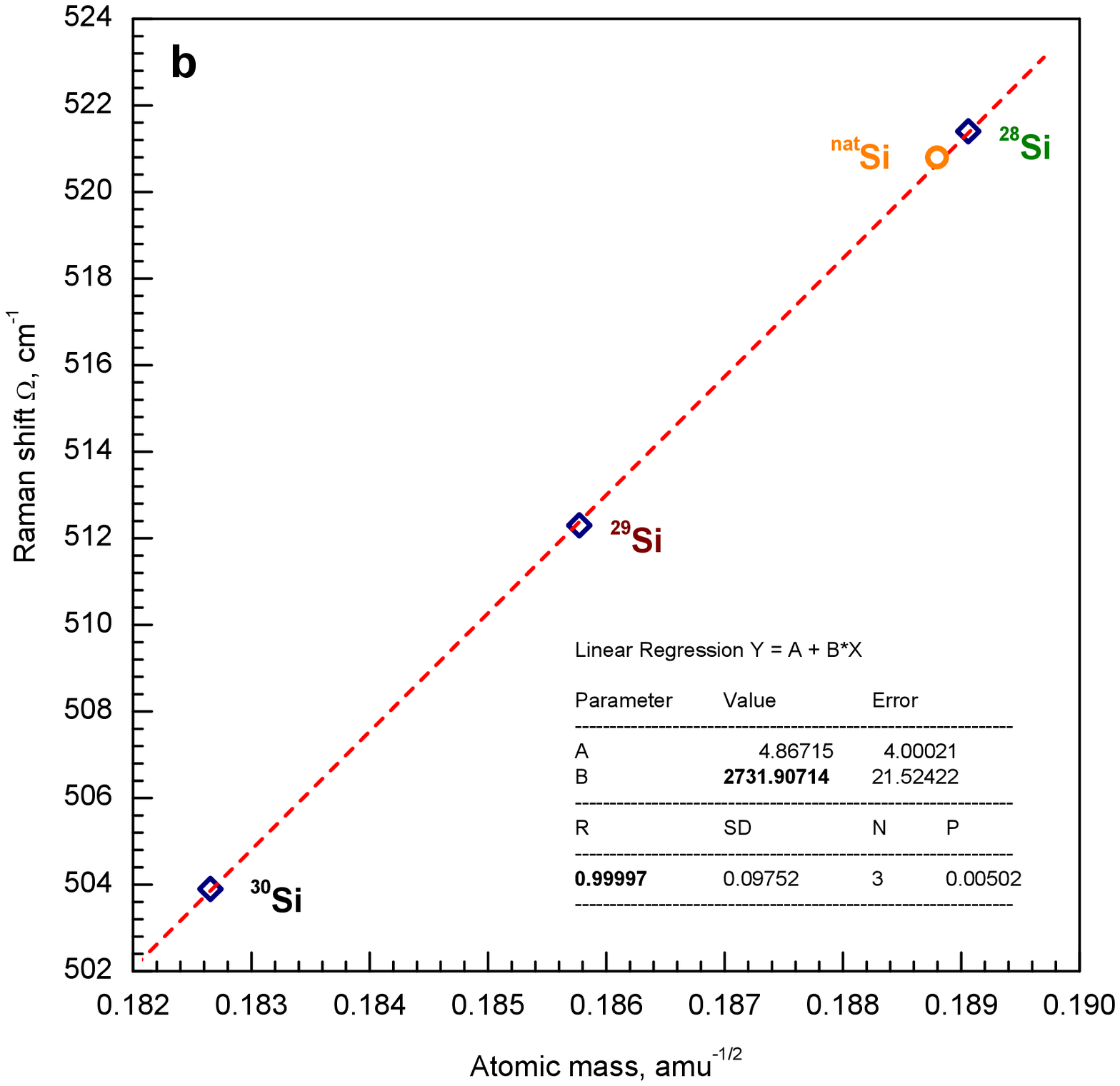}}
\caption{%
\subref{fig:1a}~Fundamental Raman band of \SiM{28}, \SiM{29}, \SiM{30}, and
\SiM{\textrm{nat}} (natural isotopic composition) single crystals. \\
\subref{fig:1b}~Peak position $\Omega$ of the Raman band vs. isotopic
crystal composition
}
\label{fig:1}
\end{figure*}
The Raman band corresponding to first-order scattering by $\Gamma_{15}$
symmetry phonons was observed in spectra of all the single crystals (showed
as circles in Fig.~\ref{fig:1a}). The band turned out to be well fitted by a
Lorentzian with width $3.3 \ldots 3.4$~\cmminusi{} for \SiM{28}, \SiM{29},
\SiM{30}{} and 3.7~\cmminusi{} for \SiM{\textrm{nat}} (solid lines in
Fig.~\ref{fig:1a}). This is an evidence of high structural perfection of
the mono-isotopic single-crystal samples.

The peak position, $\Omega$, of this Raman band is plotted vs. atomic mass,
$M_i$, in Fig.~\ref{fig:1b} for all silicon single crystals. The dashed line
represents the best fit of the experimental phonon frequency values for all the
single-crystal samples by the $\Omega = 2731.9 \left(M_i \textrm{~[a.m.u.}
\right)^{-1/2}$~\cmminusi{} dependence valid in harmonic approximation. The
root-mean-square deviation, $SD$, of experimental $\Omega$ points from the
fitted line turned out to be small for the mono-isotopic samples. Again this is
an indirect evidence of high structural perfection of those. The Raman band peak
position for \SiM{\textrm{nat}}{} sample falls slightly above the fitted line,
the band being wider than in mono-isotopic samples. This is attributable to both
anharmonicity related to different atomic masses of isotopes and to worse
structural perfection of \SiM{\textrm{nat}}{} crystal in comparison with the
mono-isotopic ones (see further).

\section{Refractive index}
\subsection{Measurements}
To measure refractive index we used the improved method of interference
refractometry \cite{14, 15, 16}. In this method firstly the order, $m$,
of the interference maximum is determined with the absolute accuracy from the
peak position, $\nu_m$, in the interference transmission spectra of
plane-parallel plates cut from the studied material, and then the refractive
index value, $n\left(\nu_m\right)$, is calculated all over the interference
transmission range measured using the expression
\begin{eqnarray}
2h \, n\left(\nu_m\right) \, \nu_m & = & m
\label{(1)}
\end{eqnarray}

The measurements were performed for plane-parallel plates of three thicknesses
(from 0.8 to 1.2~mm) cut from silicon samples of all the isotope compositions
using Bruker IFS-113v Fourier-transform vacuum spectrometer in $9500 \ldots
110$~\cmminusi{} wavenumber range ($1.05 \ldots 90$~\mkm{} wavelength range)
with 0.1~\cmminusi{} resolution. The diameter of a radiation beam transmitted
through the samples did not exceed 3~mm. At least twenty data points were
obtained for each interference maximum. The plates thickness was measured on
\mbox{DG-30}{} holographic submicrometer with the accuracy better than
0.05~\mkm. During transmission spectra and thickness measurements sample
temperature was hold to be $21.2 \pm 0.05$~\Celsius.

\begin{figure}
\includegraphics[scale=0.50,bb=50 285 545 805]{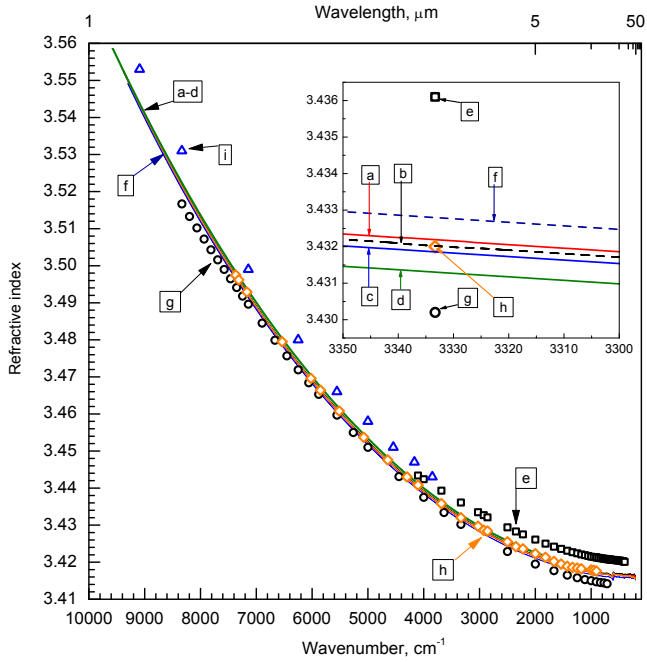}
\caption{%
Spectral dependence of the refractive index: our measurments for silicon samples
(a)~\SiM{28}, (b)~\SiM{\textrm{nat}}, (c)~\SiM{29}, (d)~\SiM{30}; the
literature data for silicon crystals with natural isotopic composition
(i)~\cite{17}, (h)~\cite{18}, (g)~\cite{19}, (e)~\cite{20}, (f)~\cite{21}.
}
\label{fig:2}
\end{figure}
Fig.~\ref{fig:2} shows the refractive index spectral dependences over the entire
measurement range for all the single crystal samples studied. As well the most
frequently cited literature data for silicon single crystals with a natural
isotopic composition \cite{17, 18, 19, 20, 21} are shown. As evident from
Fig.~\ref{fig:2}, the scatter of the refractive index data for silicon with a
natural isotopic composition exceeds considerably the difference in the
refractive index values of mono-isotopic silicon single crystals.

With silicon isotope atomic mass increasing, the refractive index spectral
dependences measured by us far from electron and phonon transition edges shifts
downwards. As an example, the refractive index values of the silicon single
crystals are listed in Table~\ref{tab:2} for 1.5, 2 and 5~\mkm{} wavelengths.
\begin{table}
\caption{%
Refractive index values of silicon single crystals at the wavelengths
1.5, 2, and 5~\mkm}
\begin{tabular}{c||D{.}{.}{5}|D{.}{.}{5}|D{.}{.}{5}}
\hline
\hline
&&&                                                 \\[-2.00ex]
Single crystal     &
\multicolumn{1}{c|}{1.5~\mkm} &
\multicolumn{1}{c|}{2~\mkm}   &
\multicolumn{1}{c}{5~\mkm}                          \\[ 0.5ex]
\hline
\hline
&&&                                                 \\[-2.00ex]
\SiM{28}           & 3.48207  & 3.45260  & 3.42194  \\
\SiM{\textrm{nat}} & 3.48191  & 3.45243  & 3.42177  \\
\SiM{29}           & 3.48171  & 3.45224  & 3.42161  \\
\SiM{30}           & 3.48112  & 3.45168  & 3.42107  \\[0.25ex]
\hline
\hline
\end{tabular}
\label{tab:2}
\end{table}

\begin{figure}
\includegraphics[scale=0.50,bb=55 115 550 810]{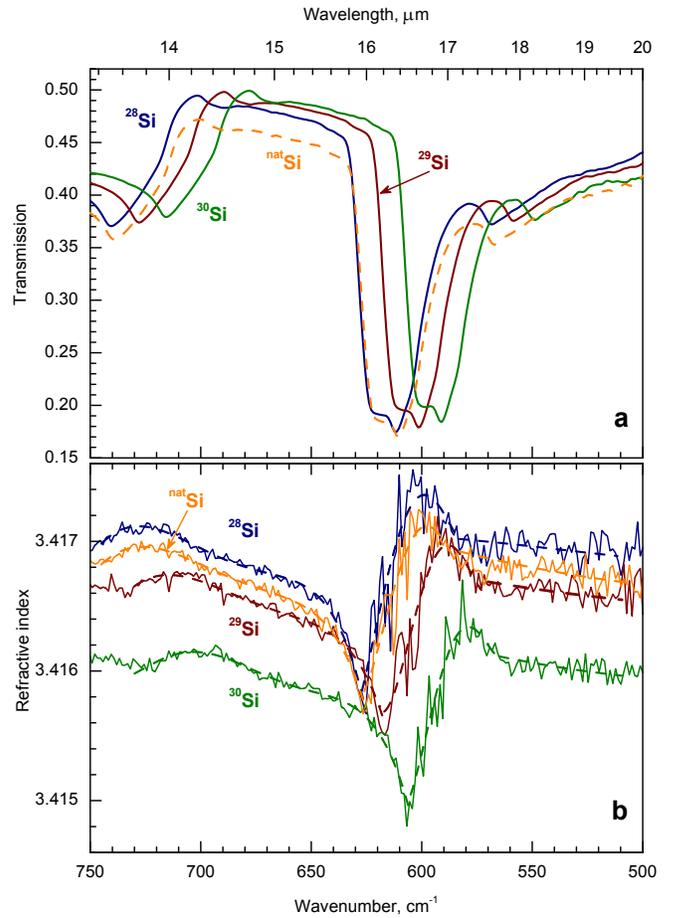}
\caption{%
(a)~Transmission spectral dependence of for 1.2-mm thick \SiM{28}, \SiM{29},
\SiM{30}, and \SiM{\textrm{nat}}  samples in the IR region of phonon absorption
peak. \\
(b)~Refractive index spectral dependence of the same single crystals.
}
\label{fig:3}
\end{figure}

It is known that two-phonon lattice absorption bands at 16.4~\mkm{}
(610~\cmminusi) with the maximum intensity of 9~\cmminusi{} are observable
in the $6.7 \ldots 20$~\mkm{} ($1500 \ldots 500$~\cmminusi) spectral range in
the spectra of $1 \ldots 2$~mm-thick silicon samples. According to
\cite{22}, those should give rise to well-distinguished features in the
refractive index dispersion. The transmission spectra measured with
2~\cmminusi{} resolution for all 1.2~mm-thick silicon crystals are shown in
Fig.~\ref{fig:3}~(a). Fig.~\ref{fig:3}~(b) presents the variation of the
crystals refractive index measured by us in this spectral region. The absorption
band shift to longer wavelengths with silicon isotope atomic mass increasing and
the anomalous behavior of refractive index dispersion well-conforming to the
classic Lorentz theory are clearly seen. As far as we know, we managed to
observe such a refractive index spectral behavior in the region of so low
absorption (less than 10~\cmminusi) for the first time. For example, in
\cite{20} (line e in Fig.~\ref{fig:2}) the interference refractometry method
was used as well to measure the refractive index of n-type silicon
(phosphorus-doped with $3 \ldots 4$~$\Omega\times\textrm{cm}$ resistivity) in
$2.43 \ldots 25$~\mkm{} wavelength range at 26~\Celsius. However, no
peculiarities in the two-phonon absorption region were revealed. In the first
approximation, the uncertainty in the refractive index values was attributed by
the authors of \cite{20} to the uncertainty in sample thickness (1 part per
$10^4$), which, in our opinion, is quite enough to observe the changes in the
spectral dependence of silicon refractive index shown in Fig.\ref{fig:3}~(b).

The refractive index values calculated from the transmission spectra are
conventionally approximated by analytical formulas allowing one to preserve the
experimental accuracy of refractive index determination. The Sellmeier or
Herzberger formulas are often used to approximate the refractive index spectral
dependence. However, our studies show that the application of generalized Cauchy
dispersion function \cite{23} allows us to reduce the root-mean-square
deviation of the approximation from the calculated refractive index values by
several times. Therefore we used it to fit the experimental data:
\begin{eqnarray}
n\left(\nu\right) & = & \sum\limits_{i=1}^{5}
\left(A_{2i} \, \nu^{2i} + B_{2i} \, \nu^{-2i} \right) + C
\label{(2)}
\end{eqnarray}
Here $\nu$ is the wavenumber, $\nu$~[\cmminusi]$ = 10^4 \times
\lambda^{-1}$~[\mkm].

\begin{figure}
\includegraphics[scale=0.50,bb=50 285 545 805]{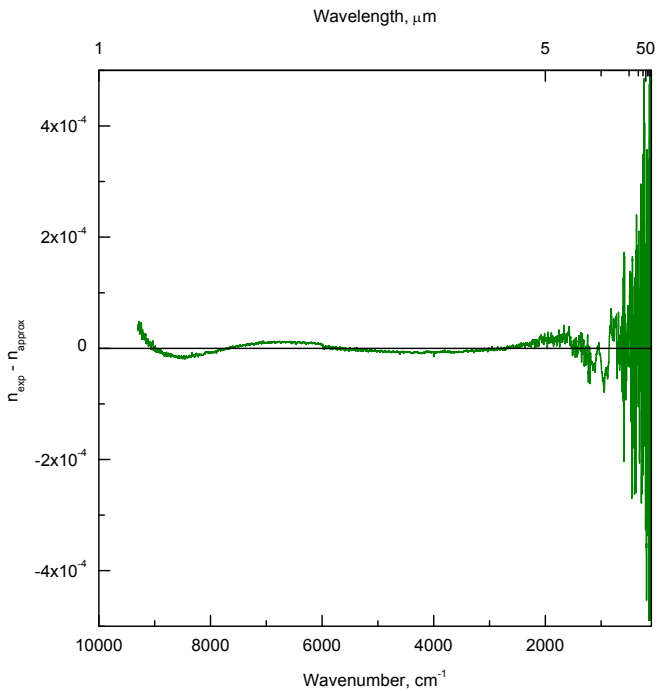}
\caption{%
Spectral dependence of the difference between the calculated, $n_\textrm{exp}$,
and approximated, $n_\textrm{approx}$, refractive index for \SiM{30} single
crystal.
}
\label{fig:4}
\end{figure}

\begin{table*}
\caption{%
Parameters of the approximation of \SiM{28}{} single crystal refractive index
}
\begin{tabular}{c||D{.}{.}{12}|D{.}{.}{12}|D{.}{.}{12}|D{.}{.}{12}}
\hline
\hline
& \multicolumn{4}{c}{Spectral range}  \\
\cline{2-5}
Parameter &
\multicolumn{1}{c|}{$9400 \ldots 745$~\cmminusi} &
\multicolumn{1}{c|}{$ 745 \ldots 625$~\cmminusi} &
\multicolumn{1}{c|}{$ 625 \ldots 583$~\cmminusi} &
\multicolumn{1}{c}{$ 583 \ldots 117$~\cmminusi} \\[-0.50ex]
&
\multicolumn{1}{c|}{($ 1.06 \ldots 13.42$~\mkm)} &
\multicolumn{1}{c|}{($13.42 \ldots 16.00$~\mkm)} &
\multicolumn{1}{c|}{($16.00 \ldots 17.15$~\mkm)} &
\multicolumn{1}{c}{($17.15 \ldots 85.47$~\mkm)} \\
\hline
\hline
&&&& \\[-2.25ex]
$A_6$ &   4.03488 \times 10^{-27} &   0.0
      &   0.0                     &  -4.47129 \times 10^{-20} \\
$A_4$ &   1.05838 \times 10^{-18} &   0.0
      &   8.07309 \times 10^{-11} &   3.89667 \times 10^{-14} \\
$A_2$ &   1.42643 \times 10^{-9}  &  -1.75666 \times 10^{-6}
      &  -1.18530 \times 10^{-4}  &  -1.00590 \times 10^{-8}  \\
$C$   &   3.41621                 &   6.74524
      &  68.44582                 &   3.41782                \\
$B_2$ & -34.48103                 &  -2.35655 \times 10^{6}
      &  -1.58031 \times 10^{7}   & -44.46299                \\
$B_4$ &   0.0                     &   7.39436 \times 10^{11}
      &   1.43557 \times 10^{12}  &   0.0                    \\
$B_6$ &   0.0                     &  -8.68186 \times 10^{16}
      &   0.0                     &   0.0                    \\[ 0.50ex]
\hline
&&&& \\[-2.00ex]
$R$   &   2.0 \times 10^{-5}      &   3.0 \times 10^{-5}
      &   9.0 \times 10^{-5}      &   1.1 \times 10^{-4}      \\[ 0.50ex]
\hline
\hline
\end{tabular}
\label{tab:3}
\end{table*}
\begin{table*}
\caption{%
Parameters of the approximation of \SiM{29}{} single crystal refractive index
}
\begin{tabular}{c||D{.}{.}{12}|D{.}{.}{12}|D{.}{.}{12}|D{.}{.}{12}}
\hline
\hline
& \multicolumn{4}{c}{Spectral range}  \\
\cline{2-5}
Parameter &
\multicolumn{1}{c|}{$9400 \ldots 730$~\cmminusi} &
\multicolumn{1}{c|}{$ 730 \ldots 617$~\cmminusi} &
\multicolumn{1}{c|}{$ 617 \ldots 573$~\cmminusi} &
\multicolumn{1}{c}{$ 573 \ldots 114$~\cmminusi} \\[-0.50ex]
&
\multicolumn{1}{c|}{($ 1.06 \ldots 13.70$~\mkm)} &
\multicolumn{1}{c|}{($13.70 \ldots 16.21$~\mkm)} &
\multicolumn{1}{c|}{($16.21 \ldots 17.45$~\mkm)} &
\multicolumn{1}{c}{($17.45 \ldots 85.72$~\mkm)} \\
\hline
\hline
&&&& \\[-2.25ex]
$A_6$ &   4.16405 \times 10^{-27} &   0.0
      &   0.0                     &   0.0                    \\
$A_4$ &   1.03238 \times 10^{-18} &   0.0
      &   1.45163 \times 10^{-10} &   4.61074 \times 10^{-15} \\
$A_2$ &   1.42694 \times 10^{-9}  &  -1.35827 \times 10^{-6}
      &  -2.03966 \times 10^{-4}  &  -5.60442 \times 10^{-10}  \\
$C$   &   3.41588                 &   5.89678
      & 110.64230                 &   3.41637                \\
$B_2$ & -29.06141                 &  -1.69322 \times 10^{6}
      &  -2.49974 \times 10^{7}   &   7.04348                \\
$B_4$ &   0.0                     &   5.12630 \times 10^{11}
      &   2.18071 \times 10^{12}  &  -7.25535 \times 10^{5}  \\
$B_6$ &   0.0                     &  -5.81255 \times 10^{16}
      &   0.0                     &   0.0                    \\[ 0.50ex]
\hline
&&&& \\[-2.00ex]
$R$   &   2.0 \times 10^{-5}      &   3.0 \times 10^{-5}
      &   1.0 \times 10^{-4}      &   1.0 \times 10^{-4}      \\[ 0.50ex]
\hline
\hline
\end{tabular}
\label{tab:4}
\end{table*}
\begin{table*}
\caption{%
Parameters of the approximation of \SiM{30}{} single crystal refractive index
}
\begin{tabular}{c||D{.}{.}{12}|D{.}{.}{12}|D{.}{.}{12}|D{.}{.}{12}}
\hline
\hline
& \multicolumn{4}{c}{Spectral range}  \\
\cline{2-5}
Parameter &
\multicolumn{1}{c|}{$9310 \ldots 720$~\cmminusi} &
\multicolumn{1}{c|}{$ 720 \ldots 605$~\cmminusi} &
\multicolumn{1}{c|}{$ 605 \ldots 565$~\cmminusi} &
\multicolumn{1}{c}{$ 565 \ldots 112$~\cmminusi} \\[-0.50ex]
&
\multicolumn{1}{c|}{($ 1.07 \ldots 13.89$~\mkm)} &
\multicolumn{1}{c|}{($13.89 \ldots 16.53$~\mkm)} &
\multicolumn{1}{c|}{($16.53 \ldots 17.70$~\mkm)} &
\multicolumn{1}{c}{($17.70 \ldots 89.29$~\mkm)} \\
\hline
\hline
&&&& \\[-2.25ex]
$A_{10}$& 0.0                     &   1.58348 \times 10^{-28}
      &   0.0                     &   0.0                    \\
$A_8$ &   0.0                     &  -3.70584 \times 10^{-22}
      &   0.0                     &   0.0                    \\
$A_6$ &   3.87178 \times 10^{-27} &   3.45410 \times 10^{-16}
      &   0.0                     &  -5.25168 \times 10^{-21} \\
$A_4$ &   1.06782 \times 10^{-18} &  -1.60314 \times 10^{-10}
      &   1.37221 \times 10^{-10} &   1.67675 \times 10^{-14} \\
$A_2$ &   1.42449 \times 10^{-9}  &   3.70631 \times 10^{-5}
      &  -1.87828 \times 10^{-4}  &  -6.42102 \times 10^{-9}  \\
$C$   &   3.41535                 &   0.0
      &  99.59448                 &   3.41676                \\
$B_2$ & -50.69159                 &   0.0
      &  -2.18377 \times 10^{7}   & -43.76842                \\
$B_4$ &   0.0                     &   0.0
      &   1.85529 \times 10^{12}  &   0.0                    \\[ 0.50ex]
\hline
&&&& \\[-2.00ex]
$R$   &   2.0 \times 10^{-5}      &   3.0 \times 10^{-5}
      &   5.0 \times 10^{-5}      &   1.0 \times 10^{-4}      \\[ 0.50ex]
\hline
\hline
\end{tabular}
\label{tab:5}
\end{table*}
\begin{table*}
\caption{%
Parameters of the approximation of \SiM{\textrm{nat}}{} single crystal
refractive index
}
\begin{tabular}{c||D{.}{.}{12}|D{.}{.}{12}|D{.}{.}{12}|D{.}{.}{12}}
\hline
\hline
& \multicolumn{4}{c}{Spectral range}  \\
\cline{2-5}
Parameter &
\multicolumn{1}{c|}{$9480 \ldots 745$~\cmminusi} &
\multicolumn{1}{c|}{$ 745 \ldots 625$~\cmminusi} &
\multicolumn{1}{c|}{$ 625 \ldots 583$~\cmminusi} &
\multicolumn{1}{c}{$ 583 \ldots 125$~\cmminusi} \\[-0.50ex]
&
\multicolumn{1}{c|}{($ 1.06 \ldots 13.42$~\mkm)} &
\multicolumn{1}{c|}{($13.42 \ldots 16.00$~\mkm)} &
\multicolumn{1}{c|}{($16.00 \ldots 17.15$~\mkm)} &
\multicolumn{1}{c}{($17.15 \ldots 80.00$~\mkm)} \\
\hline
\hline
&&&& \\[-2.25ex]
$A_6$ &   3.85272 \times 10^{-27} &   0.0
      &   0.0                     &   0.0                    \\
$A_4$ &   1.08707 \times 10^{-18} &   0.0
      &   1.06100 \times 10^{-10} &   1.10503 \times 10^{-14} \\
$A_2$ &   1.42510 \times 10^{-9}  &  -1.38069 \times 10^{-6}
      &  -1.54334 \times 10^{-4}  &  -4.28163 \times 10^{-9}  \\
$C$   &   3.41618                 &   6.01658
      &  87.37999                 &   3.41725                \\
$B_2$ & -96.96621                 &  -1.82986 \times 10^{6}
      &  -2.02501 \times 10^{7}   & -51.15422                \\
$B_4$ &   0.0                     &   5.71062 \times 10^{11}
      &   1.82698 \times 10^{12}  &  -1.93263 \times 10^{5}  \\
$B_6$ &   0.0                     &  -6.67343 \times 10^{16}
      &   0.0                     &   0.0                    \\[ 0.50ex]
\hline
&&&& \\[-2.00ex]
$R$   &   2.0 \times 10^{-5}      &   2.0 \times 10^{-5}
      &   6.0 \times 10^{-5}      &   1.3 \times 10^{-4}      \\[ 0.50ex]
\hline
\hline
\end{tabular}
\label{tab:6}
\end{table*}
Due to the presence of phonon absorption lines, the range of refractive index
approximation was subdivided into four parts. This allowed us to reduce
considerably the root-mean-square deviation between the measured and
approximated refractive index values and to bring it to the level not exceeding
$1 \times 10^{-4}$. The coefficients of Cauchy approximation (\ref{(2)}) for
refractive index of all silicon single crystals corresponding to the best fit
to the experimental results, are given in Tables~\ref{tab:3}--\ref{tab:6}. The
values of mean-square deviation for all approximation ranges are listed at the
foot of the tables. As an example, the spectral dependence of the difference
between the calculated refractive index for \SiM{30}{} single crystal and its
approximated values is shown in Fig.~\ref{fig:4}. It is seen that the deviation
of the approximation from the calculated refractive index values does not exceed
$5 \times 10^{-5}$ in a sufficiently wide range. However, with the wavelength
exceeding 16~\mkm{} the accuracy of peak position determination decreases owing
to noticeable growth of noise in the transmission spectra. Accordingly, the
accuracy of the refractive index calculation is reduced.

Comparing our results to the data published, we found that the refractive index
values closest to ours for natural silicon are given in \cite{21}. In that
work measurements were carried out by the prism method with $1 \times 10^{-4}$
relative accuracy at several wavelengths in the $1.1 \ldots 5.5$~\mkm{} range
at temperature from 20 to 300~K. The results obtained were approximated by the
Sellmeier formula.

\subsection{On the origin of the dispersion curve shift}
The observed shift of the dispersion curve is easily understood using classic
models of a lattice of oscillators with dipolar coupling \cite{24, 25}.
Neglecting the spatial dispersion the dielectric constant of a cubic lattice is
expressed in this model by the equation
\begin{eqnarray*}
\varepsilon\left(\omega\right) & = & 1 -
\frac{\alpha_i E_g\left(M_i\right)}{\omega^2 - E^2_g\left(M_i\right)} -
\frac{\beta_i  \Omega\left(M_i\right)}{\omega^2 - \Omega^2\left(M_i\right)}
\end{eqnarray*}
where $M_i$ is the atomic mass of \SiM{i} isotope, $E_g\left(M_i\right)$ is the
band gap (fundamental absorption short-wavelength edge energy),
$\Omega\left(M_i\right)$ is the phonon frequency (fundamental absorption
long-wavelength edge energy), $\alpha_i$ and $\beta_i$ are the model parameters
representing two types of oscillators corresponding to the electron and phonon
fundamental absorption edges, respectively. With no absorption, the refractive
index equals to $n\left(\omega\right) = \varepsilon^{1/2}\left(\omega\right)$.

The fundamental absorption edge energies depend on isotopic composition and for
an isotopically pure crystal can be expressed as \cite{26, 27, 28}:
\begin{widetext}
\begin{eqnarray*}
E_g\left(T,M_{i}\right) & = & E_B - a_B\left(\frac{M_{nat}}{M_i}\right)^{1/2}
\left[1 + \frac{2}{\exp\left(\Omega\left(M_{i}\right)/T\right) - 1}\right] \\
\Omega\left(M_{i}\right) & = & \Omega_0\left(\frac{M_{nat}}{M_{i}}\right)^{1/2}
\end{eqnarray*}
\end{widetext}
where $M_{nat}$ is the average atomic mass of natural silicon, $\left\{
\exp\left(\Omega\left(M_{i}\right)/T\right) - 1 \right\}^{-1}$ is phonon Bose
distribution function, $\Omega_0$ is the model parameter. According to
\cite{27}, $E_B \approx 28713.3$~\cmminusi, $a_B \approx 967.9$~\cmminusi.

The oscillator parameters, $\alpha_i$ and $\beta_i$, depend on isotopic
composition as well. First, both parameters depend on the lattice constant as
$a_0^{-3}$ and $a_0$ is known to change with isotopic composition (see e.g.
\cite{29}). However, this dependence is negligible in the qualitative model
under consideration. Second, in the case of silicon the phonon-related
parameter, $\beta_i$, describes the IR absorption due to two-phonon transitions
and hence depends on the atomic mass as $\beta_i \sim M_i^{-2}$ \cite{30}.
Therefore
\begin{eqnarray*}
\alpha_i & = & \alpha_0 \\
\beta_i  & = & \beta_0 M_i^{-2}
\end{eqnarray*}
where $\alpha_0 = \textrm{const}$, $\beta_0 = \textrm{const}$.

Thus, with Si atomic mass increasing, the electronic absorption edge shifts to
shorter wavelengths, and the phonon absorption edge shifts to longer
wavelengths. As a result, the dispersion curve of the refractive index in the
transmission window should shift downwards without considerable changes in
shape, as observed in our experiments.
\begin{figure}
\includegraphics[scale=0.50,bb=45 285 550 810]{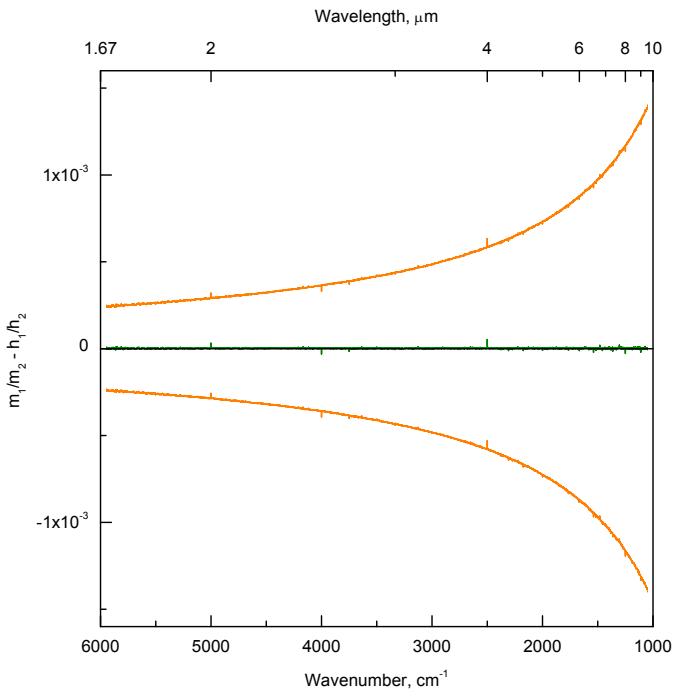}
\caption{%
Spectral dependence of the left side of equation \ref{(3)} for optimal and
differing by $\pm 1$ values of $m_{1,2}$.
}
\label{fig:5}
\end{figure}

\section{Accuracy analysis}
To measure the refractive index precisely by interference refractometry, the
orders, $m$, of interference maxima should be determined with absolute accuracy.
For this purpose we used the relation following from expression (\ref{(1)})
written for two samples of different thickness, $h_1$ and $h_2$, with
interference orders $m_1\left(\nu\right)$ and $m_2\left(\nu\right)$,
respectively:
\begin{eqnarray}
\frac{m_1\left(\nu\right)}{m_2\left(\nu\right)} -
\frac{h_1}{h_2} & = & \textrm{const}
\label{(3)}
\end{eqnarray}

If sample thickness is known with absolute accuracy, the constant in the right
side of this equation turns to zero for a certain relation between
$m_1\left(\nu\right)$ and $m_2\left(\nu\right)$. However, usually the sample
thickness can be measured with some finite accuracy resulting in non-zero value
of this constant \cite{16}. Moreover, if the transmission spectra of the
samples of different thickness are measured at different temperatures, spectral
dependence of the left side of equation (\ref{(3)}) will be affected by the
temperature dependence of the refractive index of a material. However, if one
makes a wrong choice of the maxima orders at least by 1, the spectral dependence
of the left side of equation (\ref{(3)}) will explicitly differ from constant as
evident from Fig.~\ref{fig:5} for two \SiM{\textrm{nat}}{} samples.

Minimization of the right side of dependence (\ref{(3)}) allows us to determine
uniquely the maxima orders for each pair of samples. Additionally, the
procedure of interference orders determination becomes simpler if the formula
(\ref{(1)}) is considered to be valid not only for the maxima but as well for
all other points of the spectral transmission curve for which
$m\left(\nu\right)$ values are not integer. One should use a linear
approximation for $m_{1,2}$ frequency dependence in the interval between two
nearest maxima (according to our calculations such an approximation is valid
with the accuracy about $10^{-6}$ for $1 \ldots 3$~\cmminusi{} distance between
the maxima) and increase the corresponding order by 1 passing over the next
maximum. In this case the application of the (\ref{(1)}) criterion becomes a
simple search of orders corresponding to a given frequency in the transmission
spectra of sample pairs, since the set of frequency values is the same for all
spectra.

In the method of interference refractometry the error of the refractive index
calculation depends on determination errors of all the variables occurring in
the formula (\ref{(1)}). If Fourier-transform spectrometer with the wavenumber
precision better than 0.01~\cmminusi{} is used, this error is mostly contributed
by the sample thickness measurement error, $\Delta h$. The corresponding error
in silicon refractive index is equal to $\Delta n = - n \Delta h / h$. If
$\Delta h = 0.05$~\mkm{} and values of $m$ are determined precisely for all
interference maxima, one gets $\Delta n \approx 2 \times 10^{-4}$ on average for
the only $\sim 1$~mm-thick sample used.

Using of two and more samples allows us to reduce this error due to the
correlation of the refractive index values determined for all samples over the
entire measurement range. Assuming the refractive index values to be identical
for all the samples within the entire measurement range, we can analyze the
influence of accuracy of the used thickness values on the final result of
measurements and ''correct'' these thickness values. Analysis of the
discrepancy between the ''corrected'' and initially used thicknesses provides
an idea of how well the samples are prepared and measurements are made.
Correction of thickness values ($h_i$, $i = 1, 2, 3$ in the case of three
samples) is performed by orthogonal projecting in the $(h_1, h_2, h_3)$ space
the point with coordinates equal to these thickness values on a straight line
representing solution of the following set of equations for all pairs of samples
\begin{displaymath}
\frac{h_1}{h_2} - \frac{m_1}{m_2} = 0, \quad
\frac{h_1}{h_3} - \frac{m_1}{m_3} = 0, \quad
\frac{h_2}{h_3} - \frac{m_2}{m_3} = 0
\end{displaymath}
The refractive index values scatter, $\Delta n_i$, for sample of thickness
$h_i$ is obtained immediately from the relation
\begin{displaymath}
\Delta n_i = - n_i \, \frac{\Delta h_i}{h_i}
\end{displaymath}
where $n_i$ is the ''averaged'' refractive index value, $\Delta h_i$ is the
difference between the ''corrected'' and initial thickness.

Table~\ref{tab:7} lists sample thickness measured on submicrometer at
21.2~\Celsius{} and the thickness values corrected using the interference
spectra in the middle (''MIR'', from 2 to 25~\mkm) and near (''NIR'', from 1.06
to 2~\mkm) infrared ranges, as well as the corresponding difference between the
calculated and measured thicknesses. As is seen from Table~\ref{tab:7} the
thickness correction value turns out to be extremely small for most of the
samples not exceeding the error of direct measurement of the thickness. We
estimate the average error of the refractive index determination resulting from
the error of the samples thickness determination to be $1.3 \times 10^{-4}$. The
contribution of the remaining factors influencing the samples transmission
spectra measurements and, accordingly, our calculation of the refractive index
performed similarly to paper \cite{31}, is found to be 10 times less and
therefore is negligible.
\begin{figure}
\includegraphics[scale=0.50,bb=50 285 550 810]{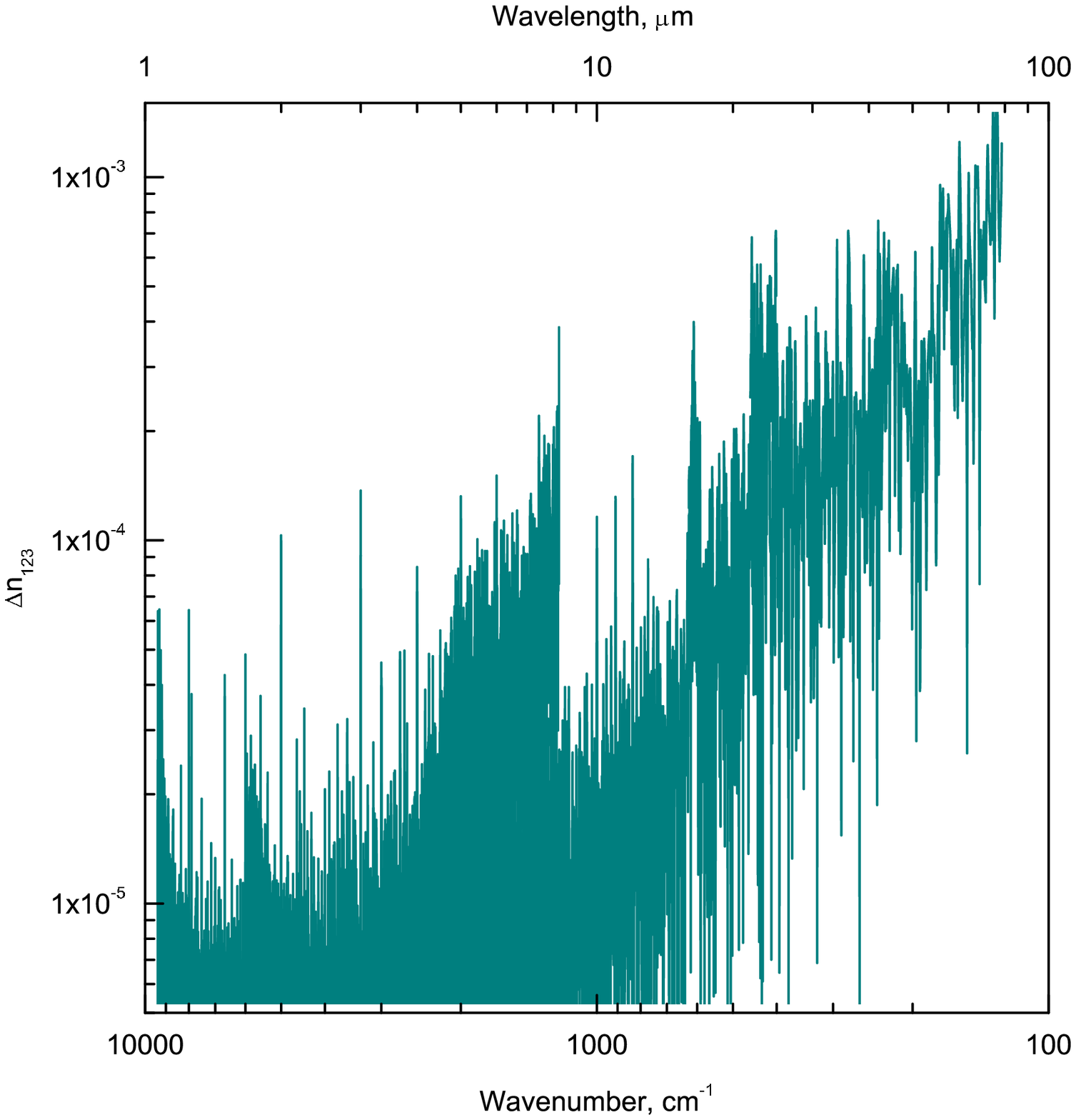}
\caption{%
Spectral dependence of the difference $\Delta n_{123}$ averaged between
refractive index for \SiM{28}{} samples of different thicknesses.
}
\label{fig:6}
\end{figure}

Unlike the errors of thickness measurement and interference maxima orders
determination, the error of the maxima positions determination is the only one
exhibiting distinct frequency dependence: the latter error grows with the
frequency decreasing (wavelength increasing) or in the absorption band vicinity
and near the transparency range edges if any in the transmission spectra. To
determine the maxima positions more exactly we restricted the range of the
sample thicknesses used and spectrometer resolution to get not less than 20
experimental points between adjacent maxima in the corresponding transmission
spectra. To estimate the influence of error of the maxima positions
determination on the refractive index calculation, we determined the averaged
difference between the refractive index values calculated for samples with
different thickness. As an example, such a difference is presented in
Fig.~\ref{fig:6} for \SiM{28}{} samples.
\begin{table}
\caption{%
Measured and corrected thicknesses of Si samples (in \mkm; $i =1, 2, 3$)
}
\begin{tabular}{c||D{.}{.}{4}|D{.}{.}{4}|D{.}{.}{4}|D{.}{.}{3}|D{.}{.}{3}}
\hline
\hline
&&&&                                                             \\[-2.00ex]
                   &
\multicolumn{1}{c|}{$h_i$ (measured)}          &
\multicolumn{1}{c|}{$h_i^\textrm{corr}$ (MIR)} &
\multicolumn{1}{c|}{$h_i^\textrm{corr}$ (NIR)} &
\multicolumn{1}{c|}{$\Delta h_{\textrm{MIR}}$} &
\multicolumn{1}{c}{$\Delta h_{\textrm{NIR}}$}                    \\[ 0.50ex]
\hline
\hline
&&&&                                                             \\[-2.00ex]
                   &  815.87 &  815.98 &  815.99 &  0.11 &  0.12 \\
\SiM{28}           & 1003.25 & 1003.23 & 1003.23 & -0.02 & -0.02 \\
                   & 1179.09 & 1179.03 & 1179.02 & -0.06 & -0.07 \\
\hline
&&&&                                                             \\[-2.00ex]
                   &  815.95 &  815.98 &  815.98 &  0.03 &  0.03 \\
\SiM{29}           & 1003.23 & 1003.21 & 1003.23 & -0.02 &  0.0  \\
                   & 1179.24 & 1179.23 & 1179.22 & -0.01 & -0.02 \\
\hline
&&&&                                                             \\[-2.00ex]
                   &  816.09 &  816.12 &  816.13 &  0.03 &  0.04 \\
\SiM{30}           & 1003.34 & 1003.41 & 1003.38 &  0.07 &  0.04 \\
                   & 1179.54 & 1179.46 & 1179.48 & -0.08 & -0.06 \\
\hline
&&&&                                                             \\[-2.00ex]
                   &  815.87 &  815.88 &  815.88 &  0.01 &  0.01 \\
\SiM{\textrm{nat}} & 1003.27 & 1003.28 & 1003.28 &  0.01 &  0.01 \\
                   & 1179.15 & 1179.14 & 1179.14 & -0.01 & -0.01 \\[ 0.50ex]
\hline
\hline
\end{tabular}
\label{tab:7}
\end{table}

\section{Summary}
In this work we have prepared and characterized high-purity single crystals of
silicon stable isotopes, \SiM{28}, \SiM{29}, and \SiM{30}{} with enrichment
above 99.9~at.\%, and of silicon with natural isotopic composition,
\SiM{\textrm{nat}}. For the first time the spectral dependences of the
refractive index of all single crystals are measured in wide transparency range
from 1.06 to about 80~\mkm{} including the phonon absorption region.

The results of our study prove the refractive index values of mono-isotopic
silicon single crystals to differ considerably. This may provide a possibility
of manufacture of multilayer wave-guiding and quantum-sized structures operating
in the IR spectrum range from different silicon isotopes to solve various
problems of electronics and fiber and integrated optics.

\begin{acknowledgments}
We are grateful to I.~D.~Kovalev and A.~M.~Potapov for determining the isotopic
and chemical compositions of isotopically enriched silicon single crystals.

This work was supported by the Russian Foundation for Basic Research (grants
No.~\mbox{08-02-00964-a} and No.~\mbox{09-03-9741-R\_Povolzh'e\_a}), by the
Russian Federation President's Grants Council for Support to the Leading
Scientific Schools of Russia (Grant No.~\mbox{NSh-4701.2008.3}) and by the
Presidium of the Russian Academy of Sciences basic research projects ''Novel
Optical Materials'' and ''Principles of Basic Research in Nanotechnologies and
Nanomaterials''.
\end{acknowledgments}

%
%

\end{document}